# BENDING DOMINATED RESPONSE OF LAYERED MECHANICAL METAMATERIALS ALTERNATING PENTAMODE LATTICES AND CONFINEMENT PLATES

**A. Amendola[1], G. Carpentieri[2], L. Feo[1], F. Fraternali[1],**

Department of Civil Engineering, University of Salerno, Via Giovanni Paolo II 132, 84084 Fisciano (SA), Italy, adaamendola1@unisa.it (A. Amendola), l.feo@unisa.it (L. Feo), f.fraternali@unisa.it (F. Fraternali)

[2]Department of Aerospace Engineering, Texas A&M University, College Station, TX 77843, USA, gerardo_carpentieri@msn.com (G. Carpentieri)

**Abstract**

*A numerical study on the elastic response of single- and multi-layer systems formed by alternating pentamode lattices and stiffening plates is presented. Finite element simulations are conducted to analyze the dependence of the effective elastic moduli of such structures upon suitable aspect ratios, which characterize the geometry of the generic pentamode layer at the micro- and macro-scale, and the lamination scheme of the layered structure. The given numerical results highlight that the examined structures exhibit bending-dominated response, and are able to achieve low values of the effective shear modulus and, contemporarily, high values of the effective compression modulus. We are lead to conclude that confined pentamode lattices can be regarded as novel metamaterials that are well suited for seismic isolation and impact protection purposes. Their elastic response can be finely tuned by playing with several geometrical and mechanical design variables.*

## 1. INTRODUCTION

Pentamode lattices are mechanical metamaterials with unconventional mechanical properties induced by the peculiar geometry of the primitive unit cell, which is formed by four rods meeting at a point. The repetition over the three dimensional space of such a cell gives rise to a diamond-like structure that supports five soft-modes of deformation (zero-energy modes), and one single rigid mode (volumetric strain) in the stretch-dominated limit [1]. Physical models of pentamode lattices have been obtained through additive manufacturing techniques at different scales, employing both metallic and polymeric materials [2][3][4].

Practical applications of pentamode metamaterials have been proposed for the realization of shear waves band-gap materials [5][6], and graded structures that make defined regions of space invisibly isolated from mechanical waves (elasto-mechanical cloak) [7][8]. More recently, pentamode lattices confined between stiffening plates have been proposed for the realization of tunable seismic isolation and impact protection devices, which show soft models controlled through the tuning of the bending moduli of members and junctions [4][9]. It has been recognized that the mechanical response of such metamaterials features some analogies with that of elastomeric bearings obtained by bonding layers of synthetic or natural rubber to stiffening plates made of steel or fiber-reinforced composites [10]-[15].

The present study aims at extending the research initiated in Refs. [4][9], through a numerical investigation on the elastic response of single- and multi-layered confined pentamode lattices featuring different aspect ratios and lamination schemes. We examine a wide range of values of selected design variables, which are related to the size of the nodal junctions (microstructure aspect ratio), the ratio between the number of unit cells placed in the vertical and horizontal directions in each layer (macrostructure aspect ratio), and the number of layers. Our goal is to extend the experimental study presented in Ref. [4] (single-layer confined pentamode lattices) to multilayer systems obtained by alternating pentamode lattices and stiffening plates. We show that a suitable design of the lattice microstructure and the lamination scheme of the examined systems leads us to obtain metamaterials featuring bending-dominated response characterized by a high ratio between the effective (uniaxial) compression modulus and the effective shear modulus. Such a result is of key importance with the aim of designing devices that have sufficiently high vertical load-carrying capacity, and contemporarily exhibit low shear rigidity. It is worth noting that many-cells, unconfined pentamode lattices do not owe such a property, exhibiting homogenized engineering constants (in the continuum limit) such that the Young and shear moduli are approximately equal each other (both are theoretically zero in the stretch-dominated limit) [16].

The structure of the paper is as follows. We begin in Sect. 2 by describing the analyzed confined pentamode lattices, and the parameters that we assume as design variables. Next, we pass to illustrate the finite element modeling that we employ to investigate on the mechanical response of such structures (Sect. 3). In Sects. 4 and 5 we present a comprehensive parametric study on the effective elastic moduli of single-layer (Sect. 4) and multi-layer (Sect. 5) confined pentamode lattices. Sect. 6 is devoted to an experimental validation of the numerical predictions of the effective elastic moduli, with reference to single-layer systems. We end in Sect. 7 by reviewing of the main results of the present study and drawing directions of future research.

## 2. LAYERED PENTAMODE LATTICES

We analyze structures composed of layers of pentamode lattices confined between stiffening plates. The examined lattices are obtained by replicating the extended face-centered-cubic (fcc) unit cell in Figure 1(a), which is formed by four primitive unit cells and 16 linkages (or rods) meeting at four distinct points (nodes/connections). Such a unit cell is periodically repeated along the $x, y, z$ axes of a Cartesian frame aligned with the unit cell edges, giving rise to layers of pentamode lattices featuring different aspect ratios (Figure 1(b)). The nodal connections are rigid and the rods are formed by the union of two truncated bi-cones featuring large diameter $D$ at the mid-span and small diameter $d$ at the extremities (size of nodal junctions) [1]-[8].

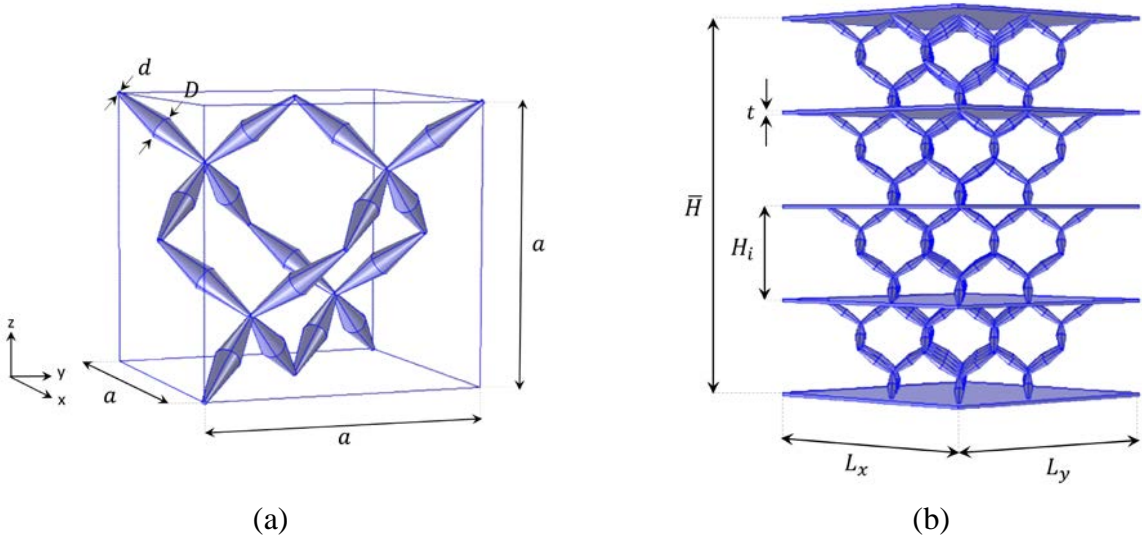

(a)          (b)

Figure 1. (a): Unit cell of pentamode lattice analyzed in the present study: extended face-centered-cubic (fcc) cell formed by rods with variable cross-section (b): Multilayer system realized obtained by alternating pentamode lattices and confinement plates.

The pentamode lattices and the stiffening plates are assumed to be made of the Ti-6Al-4V titanium alloy, which was employed in [4] to manufacture physical samples of single-layer pentamode lattices through Electron Beam Melting (EBM). As in Ref. [4], we assume $a = 30$ mm and $D = 2.72$ mm ($D/a \approx 9\%$), and we let the microstructure aspect ratio $d/a$ vary between the case with $d \approx 0$, which approximates hinged connections between the rods, and the case with $d = D$ (rigid connections). The mass density $\rho_0$, the yield strength $\sigma_{y_0}$, the Young's modulus $E_0$ and the Poisson's ratio $\upsilon_0$ of the fully-dense Ti-6Al-4V alloy are given in Table 1 [4]. We let $n_x, n_y$ and $n_z$ respectively denote the number of unit cells placed along the $x, y$ and $z$ axes in the generic layer, with $n_x = n_y$ (square pentamode layers), and $n_z = 1$ in multi-layer systems (Figure 1(b)). In addition, we let $L_x$, $L_y$ and $t = 1$ mm denote the edge lengths and the thickness of the confining plates, respectively. We also make use of the symbol $H_i$ to denote the height of the generic pentamode layer, and introduce the following notations: $H = n_z H_i$ (total height of the pentamode layers); and $A = n_x n_y a^2$ (load area). The symbol $\overline{H}$ is used to denote the overall height of the layered structure, which includes the thicknesses of the confinement plates (Figure 1(b)).

| | |
|---|---|
| $\rho_0$ [g/cm³] | 4,42 |
| $\sigma_{y_0}$ [MPa] | 910,00 |
| $E_0$ [GPa] | 120,00 |
| $\upsilon_0$ | 0,342 |

Table 1. Physical and mechanical properties of the fully dense isotropic polycrystalline Ti-6Al-4V titanium alloy forming the rods of the examined structures.

## 3. FINITE ELEMENT MODELING

We use a 3D finite element model (FEM) to study the mechanical response of the structures illustrated in the previous section. The employed FEM makes use tetrahedral solid elements to discretize both the rods of the pentamode lattices and the stiffening plates, with minimum features variable between 7% and 20% of the junction size $d$ (Figure 2). Assuming quasi-static loading conditions, we employ the MUMPS solver of COMSOL Multiphysics® to solve the linear-elastic problem of the structure under prescribed displacements of the topmost plate, by keeping the bottommost plate at rest.

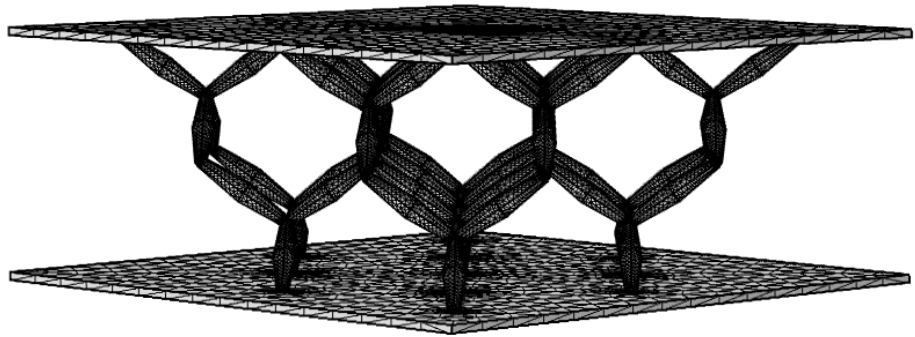

Figure 2. 3D view of a solid FEM of a confined pentamode unit.

We numerically estimate the effective shear modulus $G_c$ and the effective compression modulus $E_c$ of a laminated pentamode structure through the following formulae

$$G_c = \frac{F_h H}{\delta_h A}, \quad E_c = \frac{F_v H}{\delta_v A} \tag{1}$$

where $F_h$ and $F_v$ denote the total lateral and vertical forces measured at the top plate of the FEM, respectively under a uniform horizontal displacement $\delta_h$ (along either $x$ or $y$), and a uniform vertical displacement $\delta_v$ of the top plate.

## 4. SINGLE LAYER SYSTEMS

Figure 3 shows the variation of $E_c$ and $G_c$ with the "macroscopic" aspect ratio $H/a$ and the "microscopic" aspect ratio $d/a$, for $n_x = n_y = 2$. In such a figure and the remainder of the paper, we compare $E_c$ and $G_c$ to the Young modulus $E_r$ and the shear modulus $G_r$ of a rubber material typically employed for the manufacturing of rubber bearings ($E_r \approx 4.00$ MPa; $G_r \approx 1.00$ MPa) [17]. The results in Figure 3 show that the $E_c/E_r$ and $G_c/G_r$ ratios significantly increase with decreasing values of the $H/a$ aspect ratio (that is, in "thick" systems), especially in presence of large $d/a$ ratios (large size nodal junctions). For $H/a = 1$, we observe that it results $E_c = 0,071 E_r$ and $G_c \approx G_r/1000$ for $d/a = 0,002$; $E_c = 0,92 E_r$ and $G_c = 0,67 G_r$ for $d/a = 0,015$; $E_c = 70,17 E_r$ and $G_c = 85,26 G_r$ for $d/a = 0,9$. Since the elastic moduli of many-cells, unconfined pentamode lattices are independent of the $H/a$ ratio, and are such that the Young modulus is approximately equal to the shear modulus (both are equal to zero in the stretch-dominated limit) [16], we deduce that the above "stiffening" effects of $E_c$ and $G_c$ are due to the confinement effect played by the terminal plates against the deformation of the pentamode lattice.

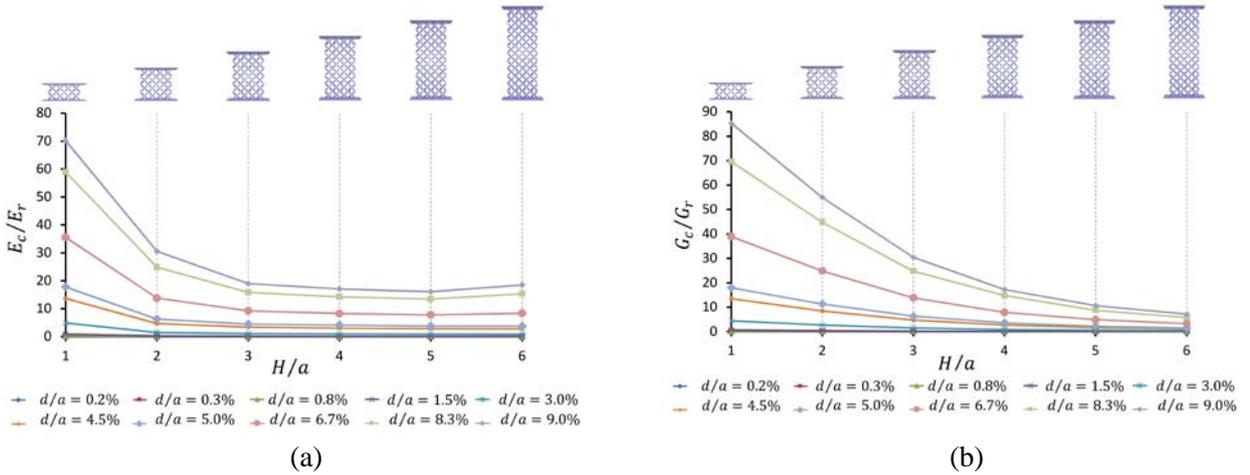

(a)                          (b)

Figure 3. Variation of the effective compression modulus $E_c$ (a) and effective shear modulus $G_c$ (b) of single-layer pentamode lattices with microscopic and macroscopic aspect ratios ($D = 2,71$ mm, $a = 15$ mm, $t = 1$ mm, $n_x = n_y = 2$).

For what specifically concerns the compression modulus $E_c$, we note that such a property is almost always larger than $E_r$, with exception to cases with $d/a < 0.015$. When it results $H/a \geq 3$, $E_c$ asymptotically tends to a constant value, for $d/a \leq 0.07$ (small size nodal junctions), or a local minimum at $H/a = 5$ ($E_c \approx 8 \div 18\, E_r$), for $d/a > 0.07$ (Figure 3(a)). The effective shear modulus $G_c$ always monotonically decreases with increasing values of $H/a$. When $H/a = 6$, it results $G_c \approx 2/10.000\, G_r$ for $d/a = 0.002$, and $G_c = 7.15\, G_r$ for $d/a = 0.09$ (Figure 3(b)).

It is worth remarking that in most applications, like, e.g., anti-seismic base isolation of buildings, one desires a sufficiently large compression modulus $E_c$ in association with a markedly low shear modulus $G_c$.

Let us now examine the variation of the $E_c/G_c$ ratio with $H/a$ and $d/a$. The results shown in Figure 4 highlight that such a ratio attains a global minimum at $H/a = 2$. For $H/a = 1$, it results: $E_c/G_c = 253.88$, $E_c/G_c = 3.29$ and $E_c/G_c = 5.48$ respectively for $d/a = 0.002$, $d/a = 0.09$, and $d/a = 0.015$. For $H/a = 4$, it instead results $E_c/G_c = 4.14$, $E_c/G_c = 3.96$ and $E_c/G_c = 4.29$ respectively for $d/a = 0.002$, $d/a = 0,09$, and $d/a = 0.015$. It is worth noting that the shear modulus of rubber-steel composite bearings is approximatively equal to that of the rubber layer ($G \approx G_r$), while the compression modulus $E_c$ of a single layer rubber-steel bearing is controlled by a shape factor $S$ defined as the ratio between the load area and the force-free (bulge) area [10]. In the case of a square rubber-steel pad with $S = 5$ (single-layer rubber thickness $t$ equal to 1/20 of the pad edge length $L$) it results $E_c \approx 169\, G_r$ [10].

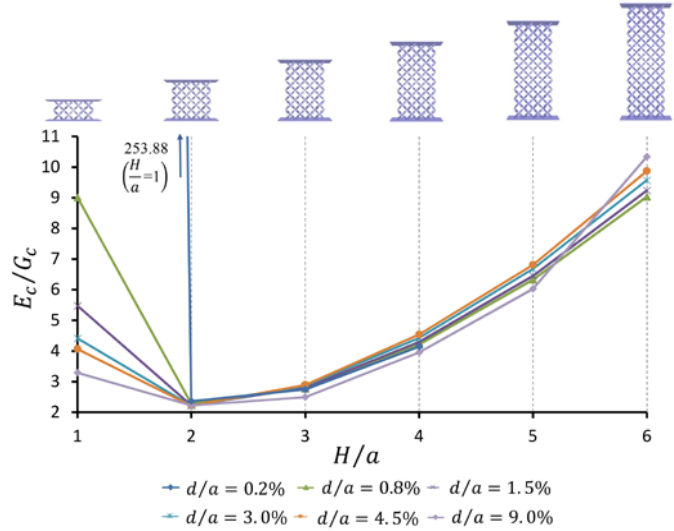

Figure 4. Variation of the $E_c/G_c$ ratio with the macro-scale ($H/a$) and micro-scale ($d/a$) aspect ratios of single-layer confined pentamode lattices ($D = 2,71$ mm, $a = 15$ mm, $t = 1$ mm, $n_x = n_y = 2$).

## 5. MULTI-LAYER SYSTEMS

We now pass to study the elastic response of multilayer systems composed of pentamode layers showing a single unit cell across the thickness ($n_z = 1$), and $2 \times 2$ unit cells in the horizontal plane ($n_x = n_y = 2$). As in the case of single layer systems, we examine lattices featuring different microscopic aspect ratios $d/a$, and various macroscopic aspect ratios $H/a$. The latter provide the number of layers forming the laminated structure, in virtue of the assumption $n_z = 1$ in each layer. The results in Figure 5 show that the distributions of the $E_c/E_r$ and $G_c/G_r$ ratios with $H/a$ and $d/a$ resemble those observed in single-layer systems (cf. Figure 3 and Figure 5).

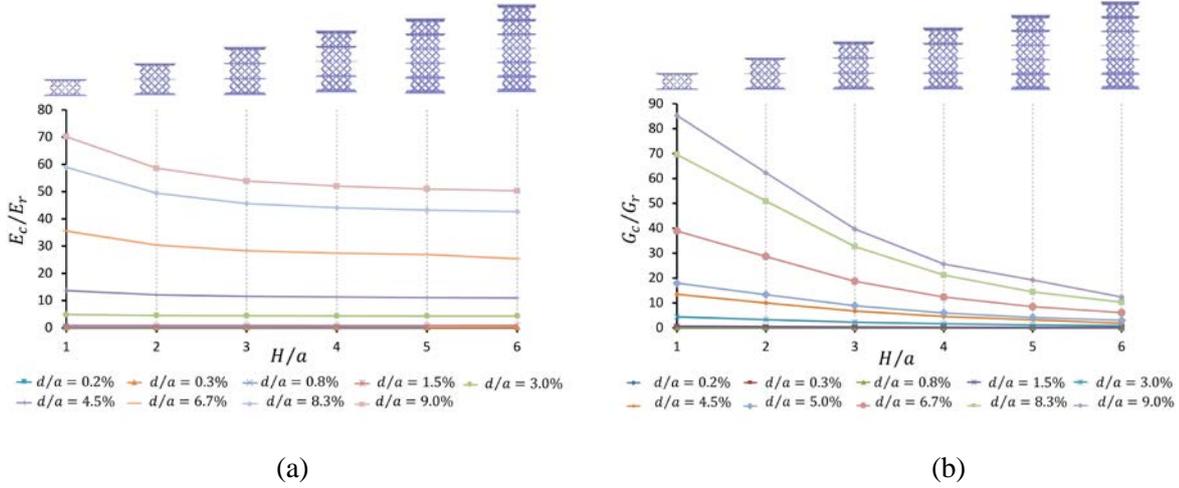

(a)                         (b)

Figure 5. Variation of the effective compression modulus $E_c$ (a) and effective shear modulus $G_c$ (b) of multi-layer pentamode lattices with microscopic and macroscopic aspect ratios ($D = 2,71$ mm, $a = 15$ mm, $t = 1$ mm, $n_x = n_y = 2$).

Figure 5 highlights that the largest values of the $E_c/E_r$ and $G_c/G_r$ ratios are obtained in the case of single layer systems ($H/a = 1$), and that the $G_c/G_r$ ratio exhibits a larger decrease rate with the number of layers, as compared to $E_c/E_r$. For $d/a = 0,002$ (small junction size), we pass from the effective moduli $E_c = 0.071 E_r$ and $G_c \approx G_r/1000$ of a single-layer structure (cf. the previous section), to the effective moduli $E_c = 0.069 E_r$ and $G_c = 1.04 \cdot 10^{-3} G_r$ of a four-layer structure ($H/a = 4$). Similarly, for $d/a = 0,09$ (large junction size), we get $E_c = 70.17 E_r$ and $G_c = 85.26 G_r$ for a single-layer structure; $E_c = 52.03 E_r$ and $G_c = 25.59 G_r$ for a four-layer structure.

We analyze the distribution of the $E_c/G_c$ ratio with $H/a$ and $d/a$ in Figure 6. The results shown in such a figure highlight that the $E_c/G_c$ ratio significantly grows with the number of layers ($H/a$), for any analyzed value of $d/a$. In the case of a five-layer structure ($H/a = 5$), we note that it results $E_c = 27.3 G_c$ for $d/a = 0.008$, and $E_c = 5.0 G_c$ for $d/a = 0.09$.

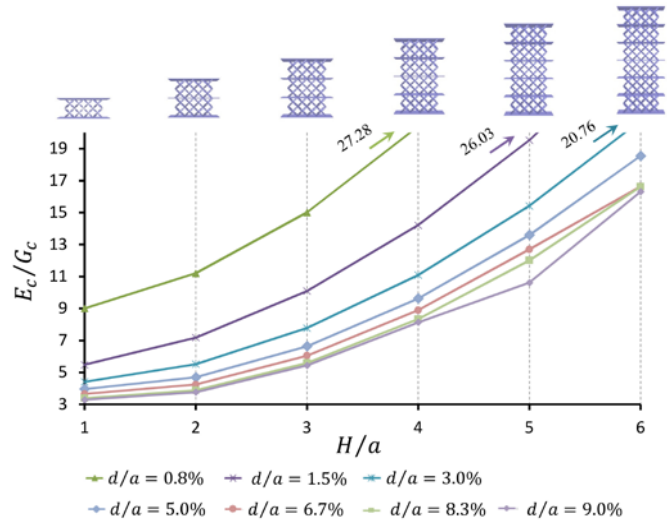

Figure 6. Variation of the $E_c/G_c$ ratio with macro-scale and micro-scale aspect ratios of multi-layer confined pentamode lattices.

Our next goal is to study the dependence of the $E_c/E_0$ and $G_c/G_0$ ratios on the solid volume fraction of the unit cell $\phi$ (volume of the rods in the unit cell divided by the volume of the unit cell). Such a study is aimed at detecting the nature of the overall response of the confined structure (stretching-dominated or bending-dominated) [18]-[20]. The $E_c/E_0$ vs. $\phi$ and $G_c/G_0$ vs. $\phi$ plots in Figure 7(a,b) refer to systems featuring different values of $n_x \times n_y \times n_l$, where $n_l$ denotes the number of layers. It is seen that the $E_c/E_0$ and $G_c/G_0$ ratios vary with $\phi$ according to non-linear laws, which proves evidence of a bending-dominated response of the examined structures [18]-[20] (quadratic regressions of the FEM results are represented by solid lines in Figure 5).

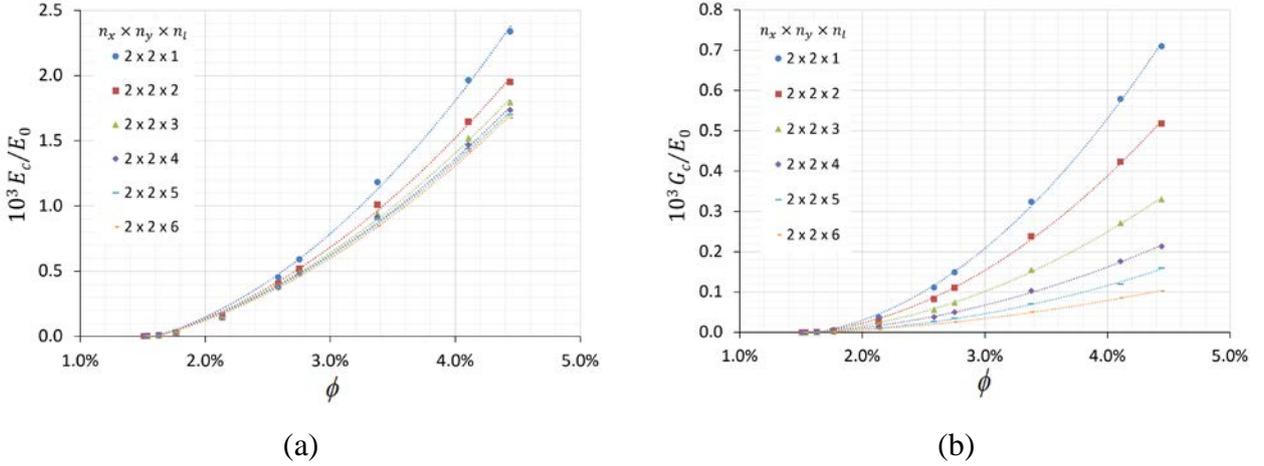

(a) (b)

Figure 7. Variation of the $E_c/E_0$ and $G_c/G_0$ ratios with the solid volume fraction $\phi$ (markers indicate FEM results; solid lines denote quadratic regressions of FEM results).

We conclude our parametric study by comparing the $E_c/G_c$ ratios of single- and multi-layer structures at constant $H/a = 4$, for different values of the microstructure aspect ratio $d/a$ (Figure 8). The results in Figure 8 highlight that a four-layer structure with small junction size ($d/a = 0.002$) achieves the ratio $E_c/G_c \approx 267$, which is greater than the $E_c/G_c$ ratio of a square rubber pad confined by rigid plates with thickness equal to 1/20 of the edge length ($E_c/G_c \approx 169$, cf. the previous section).

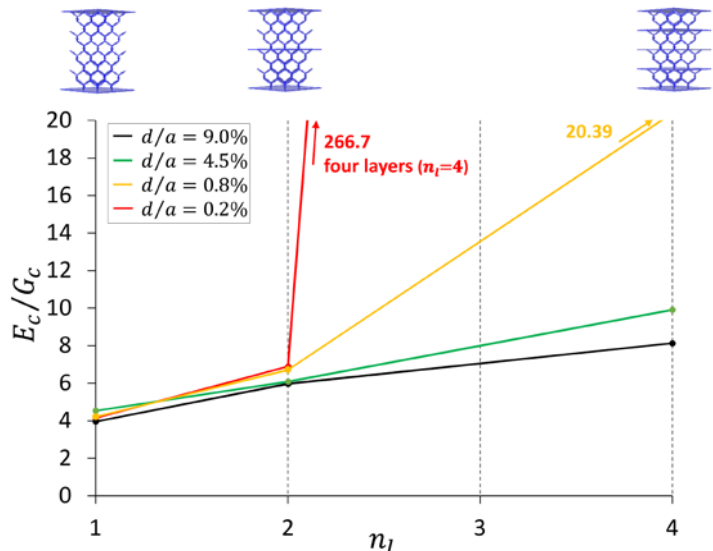

Figure 8. Comparison of the $E_c/G_c$ ratios of single- and multi-layer structures.

## 6. EXPERIMENTAL VALIDATION

The current section presents an experimental validation of the solid FEM described in Sect. 3 against the results of quasi-static laboratory tests on EBM samples of pentamode structures [4]. Such a validation relates experimental and numerical values of vertical and horizontal stiffness properties of single-layer pentamode lattices featuring thick and slender macroscopic aspect ratios.

The analyzed lattices are composed of two unit cells in the horizontal plane ($n_x = n_y = 2$) and varying number of unit cells along the $z$-axis. Following the notation given in [4], we hereafter name TPM the systems featuring $n_z = 2$ (*"thick pentamode materials"*), and SPM systems featuring $n_z = 4$ (*"slender pentamode materials"*). For each of such systems, we analyze physical samples featuring different values of $d$ ($d_1, d_2, d_3$) and fixed values of $a$ and $D$, as shown in Table 2. We use the label *i* to denote the SPM /TMP sample featuring $d = d_i$.

|  | $a$ [mm] | $D$ [mm] | $d_1$ [mm] | $d_2$ [mm] | $d_3$ [mm] |
|---|---|---|---|---|---|
| Built size | 30 | 2.72 | 0.49 | 1.04 | 1.43 |
| (CAD size) | (30) | (2.71) | (0.45) | (0.90) | (1.35) |

Table 2. Geometrical properties of physical samples of confined pentamode lattice analyzed in Ref [4].

SPM and TPM specimens have been subject to quasi-static shear and compression tests in Ref. [4]. Shear tests consisted of cyclic lateral force ($F_h$) vs. lateral displacement ($\delta_h$) tests in displacement control, under constant applied vertical load. Compression tests instead consisted of vertical force ($F_v$) vs. vertical displacement ($\delta_v$) tests under zero applied lateral force. Hereafter we let $\overline{K}_{h,eff}$ and $\overline{K}_{v,eff}$ denote the mean values of the effective (secant) stiffness coefficients experimentally determined as described in [4]. Such coefficients are used to determine experimental values of the effective elastic shear modulus $G_c$ and the effective elastic compression modulus $E_c$, through the following equations

$$G_{c,exp} = \overline{K}_{h,eff} \cdot \frac{H}{A}, \qquad E_{c,exp} = \overline{K}_{v,eff} \cdot \frac{H}{A} \qquad (2)$$

Let us now compare the experimental effective moduli with finite element predictions $G_{c,fem}$ and $E_{c,fem}$ of the same quantities (Table 3).

|  | SPM1 | SPM2 | SPM3 | TPM1 | TPM2 | TPM3 |
|---|---|---|---|---|---|---|
| $G_{c,exp}$ | 70,43 | 444,77 | 1.033,01 | 222,04 | 1.061,90 | 1.957,77 |
| $G_{c,fem}$ | 118,20 | 860,73 | 2.702,73 | 372,63 | 2.685,00 | 8.441,83 |
| $E_{c,exp}$ | 501,49 | 3.384,47 | 8.165,70 | 657,98 | 2.971,50 | 8.538,12 |
| $E_{c,fem}$ | 507,67 | 4.811,00 | 12.265,67 | 829,18 | 5.935,67 | 18.69,67 |

Table 3. Experimental and FEM values of the effective shear modulus and effective compression modulus of SPM and TPM specimens (kPa).

The results in Table 3 show a good agreement between the orders of magnitude of experimental and numerical moduli. It is seen that such quantities markedly increase for increasing values of the size $d$ of the nodal junctions, which is related to the bending rigidity of the lattice (as we noticed before, perfect pin joints are obtained in the limit $d/a \to 0$, while the case with $d/a > 0$ corresponds to nonzero bending rigidities of nodal junctions and rods). The numerical predictions of the effective moduli are always larger than the experimental values, both in SPM and TPM systems. Such a result is explained by approximation errors of the finite element solutions, and the fact that such simulations do not capture the micro-plasticity damage that was observed in the experiments before macro-yielding [4],[21]. By examining the results in Table 3, we find out that the $G_{c,fem}/G_{c,\exp}$ ratio is equal to 1,69 (1,67) in SPM1 (TPM1); 1,94 (2,52) in SPM2 (TPM2); and 2, 90 (4,31) in SPM3 (TPM3). The $E_{c,fem}/E_{c,\exp}$ ratio is instead equal to 1,01 (1,26) in SPM1 (TPM1); 1,42 (2,00) in SPM2 (TPM2); and 1,50 (2,19) in SPM3 (TPM3). The growth of the FEM approximation errors with $d$ follows by the increased role played by micro-plasticity damage effects on the experimental response of SPM and TPM specimens, as the size of the nodal junctions gets larger [4]. It is worth noting that both experimental results and finite element simulations lead to (effective) compression moduli greater than shear moduli, for each analyzed system.

## 7. CONCLUDING REMARKS

We have conducted a numerical study on the elastic response of single-layer and multi-layer systems obtained by alternating pentamode lattices and stiffening plates. The elastic response of such systems turned out to be different from that of unconfined pentamode lattices, due to the confinement effect played by the stiffening plates. While in unconfined pentamode lattices the Young modulus is approximately equal to the shear modulus [16], in confined pentamode lattices we observe that the effective compression modulus $E_c$ is always larger than the effective shear modulus $G_c$. In the case of single-layer systems, the results given in Sect. 4 highlight that the $E_c/G_c$ ratio attains a global minimum at $H/a = 2$ ($E_c/G_c \approx 2$), and significantly grows both towards $H/a = 1$ (thick systems), and for $H >> a$ (slender systems). For multi-layer systems, on the other hand, the $E_c/G_c$ ratio monotonically increases with the number of layers (cf. Sect. 5). The validation of numerical predictions of the effective moduli against the experimental data presented in Ref. [4] highlighted good theory vs. experiment matching, and confirmed the result that the compression modulus of a confined pentamode lattice is always greater than the shear modulus (Sect. 7).

Overall, the results of the present study highlight several analogies between the response of confined pentamode lattices and that of rubber bearings alternating layers of rubber and stiffening plates, which are commonly employed as seismic isolation devices [10]-[15]. In fact, the role played by the stiffening plates is similar in such systems, being mainly devoted to stiffen the vertical deformation mode of the structue. The use of confined pentamode lattices as novel impact protection devices and seismic isolators ("pentamode bearings") deserves special attention, based on the following considerations:

- the mechanical properties of the pentamode layers forming such devices mainly depend on the geometry of the lattice, more than on the chemical nature of the employed materials (metallic, ceramic, polymeric, etc.);
- it is easy to adjust the mechanical properties of pentamode bearings to those of the structure to be protected/isolated, by playing with the lattice geometry and the nature of the material, as opposed to rubber bearings, where instead the achievement of very low shear moduli implies marked reductions of the vertical load carrying capacity, making such devices not particularly convenient in the case of structures with very high fundamental periods of vibrations (such as, e.g., very tall buildings; highly compliant structures; very soft soils; etc.) [12],[22]-[23];

- the dissipation of pentamode bearings can be conveniently designed through an accurate choice of the material to be used for the pentamode lattices, and inserting, - when necessary, an additional dissipative element within the device (such as, e.g., a lead core);

- the possibility to design and fabricate laminated composite bearings showing layers with different materials, geometries and properties: such a design approach is instead much less effective in the state-of-the-art laminated rubber bearings, where the only lamination variable consists of the type of rubber to be employed for the soft pads (natural rubber or synthetic rubber);

- the freedom in the choice of the materials of the pentamode lattices, by keeping the elastic properties of the device essentially unchanged, allows the designer to adapt the energy dissipation capacity and the life span (i.e., the durability) of the device to the actual use conditions [22]-[23];

- the possibility to replace the fluid components of the structural bearings and energy absorbing devices currently available on the market (such as, e.g., viscous fluid dampers and tuned mass dampers) with pentamode lattices: such a replacement would lead to significantly reduce the technical issues related to fluid leaking and frequent maintenance, which currently affect the state-of-the-art devices involving fluid materials;

- the mechanical properties of pentamode bearings can be dynamically adjusted and measured, by equipping selected struts of the pentamode lattices with sensors and/or actuators;

- pentamode bearings can be directly manufactured from computer-aided design data outputted by a computational material design phase, on employing advanced and fast additive manufacturing techniques at different scales, and single or multiple materials (metals, polymers, etc.).

Several aspects of the present work pave the way to relevant further investigations and generalizations that we address to future work. First, mechanical models for composite rubber-steel bearings [10] need to be generalized to pentamode-steel bearings, accounting for the peculiar deformation models of such systems, and discrete-to-continuum approaches [16],[24]-[25]. Second, physical models of pentamode isolators need to be constructed, employing, e.g., additive manufacturing techniques [2]-[4] or manual assembling methods [26]. An experimental verification phase is also needed to assess the actual isolation and dissipation capabilities of pentamode bearings arising, e.g., from inelastic response and/or material fracture [27][28]. Another relevant generalization of the present research regards the design of dynamically tunable systems based on the insertion of structural hinges [29] and/or prestressed cables within pentamode lattices, with the aim of designing novel metamaterials and bio-inspired lattices tunable by nodal stiffness properties (semi-rigid nodes; dissipative junctions, etc.), as well as local and global prestress [30].

## Acknowledgements

The authors are grateful to Mariella De Piano (Department of Civil Engineering, University of Salerno) for helpful assistance with finite element simulations, and Gianmario Benzoni (Department of Structural Engineering, University of California, San Diego) for guidance and assistance with the experimental aspects of the present research.

## REFERENCES

[1] Milton GW, Cherkaev AV. Which Elasticity Tensors are Realizable? J Eng Mater Technol 1995; 117(4): 483-493.

[2] Kadic M, Bückmann T, Stenger N, Thiel M, Wegener M. On the practicability of pentamode mechanical metamaterials. Appl Phys Lett 2012; **100**:191901.

[3] Schittny M, Bückmann T, Kadic M, Wegener M. Elastic measurements on macroscopic three-dimensional pentamode metamaterials. Appl Phys Lett 2013;103:231905,.

[4] Amendola A, Smith CJ, Goodall R, Auricchio F, Feo L, Benzoni G, Fraternali F. Experimental response of additively manufactured metallic pentamode materials confined between stiffening plates. Compos Struct 2016;142:254-262.

[5] Martin A, Kadic M, Schittny R, Bückmann T, Wegener M. Phonon band structures of three-dimensional pentamode metamaterials. Phys Rev B 2010;86:155116.

[6] Huang Y, Lu X, Liang G, Xu Z. Pentamodal property and acoustic band gaps of pentamode metamaterials with different cross-section shapes. Phys Lett A 2016;380(13):1334-1338.

[7] Bückmann T, Thiel M, Kadic M, Schittny R, Wegener M. An elastomechanical unfeelability cloak made of pentamode metamaterials. Nat Comm 2014;5:4130.

[8] Chen Y, Liu X, Hu G. Latticed pentamode acoustic cloak. Scientific Reports 2015, 5:15745.

[9] Fraternali F, Carpentieri G, Montuori R, Amendola A, Benzoni G. On the use of mechanical metamaterials for innovative seismic isolation systems, COMPDYN 2015 - 5th ECCOMAS Thematic Conference on Computational Methods in Structural Dynamics and Earthquake Engineering, 349-358.

[10] Kelly JM. Earthquake-resistant design with rubber. London: Springer-Verlag, 1993.

[11] Benzoni G, Casarotti C. Effects of vertical load, strain rate and cycling on the response of lead-rubber seismic isolators. J Earthquake Eng 2009;13(3):293-312.

[12] Higashino M, Hamaguchi H, Minewaki S, Aizawa S. Basic characteristics and durability of low-friction sliding bearings for base isolation. Earthquake Engineering and Engineering Seismology 2003;4(1):95-105.

[13] Toopchi-Nezhad H, Tait MJ, Drysdale RG. Bonded versus unbonded strip fiber reinforced elastomeric isolators: finite element analysis. Compos Struct 2014;93(2):850-9.

[14] Osgooei PM, Tait MJ, Konstantinidis D. Three-dimensional finite element analysis of circular fiber-reinforced elastomeric bearings under compression. Compos Struct 2014;108(1):191–204.

[15] Osgooei PM, Tait MJ, Konstantinidis D. Finite element analysis of unbonded square fiber-reinforced elastomeric isolators (FREIs) under lateral loading in different directions. Compos Struct 2014;113(1):164-73.

[16] Norris AN. Mechanics of elastic networks. Proceedings of the Royal Society of London A 2014; 470:20130611.

[17] Gere JM, Timoshenko, SP. Mechanics of Materials. 5th ed. Boston: PWS Kent Publishing, 1970.

[18] Meza LR, Das S, Greer JR. Strong, light weight, and recoverable three-dimensional ceramic nanolattices. Science 2014;6202(345):1322-1326.

[19] Deshpande VS, Fleck NA, Ashby MF. Effective properties of the octet-truss lattice material. J Mech Phys Solids 2001;22:409-428.

[20] Schaedler TA, Jacobsen AJ, Torrents A, Sorensen AE, Lian J, Greer JR, Valdevit L, Carter WB. Ultralight Metallic Microlattices. Science 2011;6058(334):962-965.

[21] Steele RK, McEvily AJ. The high-cycle fatigue behavior of Ti-6Al-4V alloy. Eng Fract Mech 1976;8:31-37.

[22] Monaldo-Mercaldo JC. Passive and Active Control of Structures. MSc Thesis in Civil and Environmental Engineering, Massachusetts Institute of Technology, USA, 1995.

[23] Oladimeji FA. Bridge Bearings - Merits, Demerits, Practical Issues, Maintenance and Extensive Surveys on Bridge Bearings, MSc Thesis, Royal Institute of Technology (KTH), Department of Civil and Architectural Engineering, Division of Structural Engineering and Bridges, Stockholm, Sweden, 2012.

[24] Gurtner G, Durand M. Stiffest elastic networks. Proceedings of the Royal Society of London A 2014; 470:20140522.

[25] Schmidt B, Fraternali F, Universal formulae for the limiting elastic energy of membrane networks. J Mech Phys Solids 2012;60:172-180.

[26] Amendola A, Carpentieri G, de Oliveira M, Skelton RE, Fraternali F. Experimental investigation of the softening stiffening response of tensegrity prisms under compressive loading. Compos Struct 2014;117:234-243.

[27] Modano M, Fabbrocino F, Gesualdo A, Matrone G, Farina I, Fraternali F. On the forced vibration test by vibrodyne. In: COMPDYN 2015 Conference Proceedings, 25-27 May 2015.

[28] Fraternali F. Free Discontinuity Finite element models in two-dimensions for in-plane crack problems. Theor Appl Fract Mec 2007;47:274-282.

[29] Chilton J. Space Grid Structures, Oxford, UK, 2000.

[30] Fraternali F, Carpentieri G, Amendola A, Skelton RE, Nesterenko VF. Multiscale tunability of solitary wave dynamics in tensegrity metamaterials. Appl Phys Lett 2014;105:201903.